# Tetragonal superstructure of the antiskyrmion hosting Heusler compound Mn$_{1.4}$PtSn

Praveen Vir, Nitesh Kumar, Horst Borrmann, Bayardulam Jamijansuren, Guido Kreiner, Chandra Shekhar, and Claudia Felser*

Max Planck Institute for Chemical Physics of Solids 01187 Dresden, Germany

**ABSTRACT:** Skyrmions in non-centrosymmetric magnets are vortex-like spin arrangements, viewed as potential candidates for information storage devices. The crystal structure and non-collinear magnetic structure together with magnetic and spin-orbit interactions define the symmetry of the Skyrmion structure. We outline the importance of these parameters in the Heusler compound Mn$_{1.4}$PtSn which hosts antiskyrmions, a vortex-like spin texture related to skyrmions.[1] We overcome the challenge of growing large micro-twin-free single crystals of Mn$_{1.4}$PtSn which has proved to be the bottleneck for realizing bulk skyrmionic/antiskyrmionic states in a compound. The use of 5d-transition metal, platinum, together with manganese as constituents in the Heusler compound such as Mn$_{1.4}$PtSn is a precondition for the non-collinear magnetic structure. Due to the tetragonal inverse Heusler structure, Mn$_{1.4}$PtSn exhibits large magneto-crystalline anisotropy and $D_{2d}$ symmetry, which are necessary for antiskyrmions. The superstructure in Mn$_{1.4}$PtSn is induced by Mn-vacancies which enables a ferromagnetic exchange interaction to occur. Mn$_{1.4}$PtSn, the first known tetragonal Heusler superstructure compound, opens up a new research direction for properties related to the superstructure in a family containing thousands of compounds.

Magnetic skyrmion is an ideal candidate for efficient future memory devices because it can be driven using an ultra-low current density.[2] Spin textures in skyrmions are topologically stable and can therefore also lead to interesting physical properties, such as the topological Hall effect and the generation of spin transfer torque.[3,4] The vast body of work in this field has established essential ingredients for the stability of skyrmions.[5,6] However, to come out with a recipe in the form of a compound, which contains all these essential ingredients is a difficult task. The most important criteria to stabilize a skyrmion in a magnet is its crystal structure should lack inversion symmetry. The induced Dzyaloshinskii–Moriya interaction (DMI) produced in this type of non-centrosymmetric compound ensures that the spins are aligned in a non-collinear fashion.[7,8] Secondly, the stability of skyrmions at technologically relevant temperatures requires a large exchange interaction. The magnetic to non-magnetic transition temperature ($T_C$) in large exchange interaction systems is usually high. Another favorable parameter for the stabilization of skyrmions over a broad range of temperature and magnetic field strength is the availability of easy axis, which can be attained in tetragonal or hexagonal crystal systems with large spin-orbit coupling.[9]

Herein, we present single crystals of the antiskyrmion hosting[1] Heusler compound, $Mn_{1.4}PtSn$, wherein all the design principles mentioned above are ideally applied. Taking the example of $Mn_{1.4}PtSn$, the main goal of this work is to demonstrate how the chemical and structural tuning in a material be carried out to come up with potential candidates for skyrmions/antiskyrmions. The structure of $Mn_{1.4}PtSn$ originates from the inverse Heusler compounds and therefore, it does not contain any center of inversion. The large spin-orbit coupling associated with the heavy elements Pt and Sn in this compound further enhances the DMI and magnetocrystalline anisotropy effects. A high $T_C$ of ~400 K in tetragonally distorted $Mn_{1.4}PtSn$ indicates its strong exchange interaction. The tetragonal structure of $Mn_{1.4}PtSn$ is similar to many Mn-rich Heusler compounds, where Jahn–Teller-like distortion and spin-orbit coupling play important roles. Interestingly, the tetragonal structure of inverse Heusler compounds exhibits a special $D_{2d}$ symmetry, which provides all the structural ingredients to realize another vortex-like spin texture, known as an antiskyrmion, which is topologically related to the skyrmions.[1,10-12]

However, the biggest challenge in $Mn_{1.4}PtSn$ and its related compounds is the formation of single crystals, which is complicated by a high temperature cubic austenite to tetragonal martensitic structural transition, a well-known phenomenon for shape memory alloys.[13-15] Therefore, conventional crystal growth techniques always result in micro-twinned crystals. In this manuscript we show that under certain reaction conditions large micro-twin-free single crystals of $Mn_{1.4}PtSn$ can be grown, which is essential for establishing the intrinsic features of antiskyrmions in this compound. Micro-twin-free single crystals allow us to determine the structure of $Mn_{1.4}PtSn$ using single crystal X-ray diffraction, which lead to the discovery of the first known superstructure of a Heusler compound originating from Mn-vacancy ordering. The additional tetragonality due to the formation of the superstructure can have favorable effects for the stability of antiskyrmions in $Mn_{1.4}PtSn$ and many other related compounds.

Heusler compounds are a promising class of materials because they demonstrate various exciting physical properties, e.g., superconductivity, thermoelectricity, hard magnetism, magnetic semiconductors, and Berry curvature induced topology. One of the main advantages of these materials is that their properties can easily be tuned simply by varying their valence electrons. This class of compounds have empirical formula XYZ (half Heusler) and $X_2YZ$ (full Heusler), where X and Y are transition metals and Z is a main group element.[16] Full Heusler compound $X_2YZ$ can be of two types; regular Heusler and inverse Heusler. Normally, the compounds crystallize in cubic structures except when the Y or/and Z elements are heavy metals with large spin-orbit coupling, in which case the crystal field effect becomes large enough to induce a tetragonal distortion. The regular cubic Heusler structure ($Fm\bar{3}m$), and the inverse cubic Heusler structure ($F\bar{4}3m$), therefore, transform into tetragonal Heusler structures having space groups $I4/mmm$ and $I\bar{4}m2$, respectively.[16-19]

The formation of the desired compound $Mn_{1.4}PtSn$ is highly sensitive to the initial ratios of Mn to Pt during the crystal growth. The initial ratios of 1.4-2:1 for Mn and Pt, afford single crystals of an undesired compound having the composition, $Mn_{22}Pt_{30}Sn_{48}$. To suppress the formation of $Mn_{22}Pt_{30}Sn_{48}$, we took the relatively high Mn and Pt ratio of 3:1, which resulted in the formation of $Mn_{1.4}PtSn$. During the flux growth, when the reaction mixture was cooled from high temperatures (~1273 K), the obtained crystals were always micro-twinned. The formation of micro-twinning can be understood in terms of martensitic phase transition, wherein single crystals are initially formed in the cubic phase, after which they transform into the tetragonal phase upon cooling, affording micro-twinned plates of different orientations.

To avoid the formation of micro-twins, we performed differential scanning calorimetry (DSC) on a polycrystalline $Mn_{1.4}PtSn$ sample prepared by arc-melting, as shown in **Figure** S1(b). Other than a sharp peak at 1306 K, which corresponds to the melting point of the compound, we observe another peak at 1013 K (inset in **Figure** S1(b)), which is related to the structural transition from the high-temperature cubic phase to the low-temperature tetragonal phase. Considering the structural transformation at 1013 K, we have chosen the crystal growth range below this temperature. First, we set the furnace temperature to 1323 K, after which rapid cooling to 923 K occurred (below the structural transformation temperature), to avoid the growth of any cubic phase. The slow cooling from 923 to 723 K resulted in the formation of single crystals free of micro-twins. Methods such as Bridgman crystal growth and optical floating zone require direct cooling from the melt; consequently, crystals of $Mn_{1.4}PtSn$ from these methods are always micro-twinned. Therefore, the only successful method to grow crystals of $Mn_{1.4}PtSn$, free of micro-twins, is the flux-growth, wherein the growth process can be controlled far below the melting point. Stoichiometric $Mn_2PtSn$ is supposed to crystallize in the space group $I\bar{4}m2$, but this stoichiometric compound cannot be synthesized in a bulk single phase. However, removing some manganese atoms from the unit cell leads to the stabilization of the single-phase compound, $Mn_{1.4}PtSn$. It crystallizes in a different space group ($I\bar{4}2d$), which is a superstructure of statistically vacancy distributed $Mn_{1.4}PtSn$, obtained from $Mn_2PtSn$ (space group $I\bar{4}m2$); it will be discussed later in this article.

In Figure 1, we demonstrate how the structural evolution occurs from an inverse cubic Heusler structure to an inverse tetragonal Heusler structure, followed the vacancies formation and their ordering, which transforms it into the superstructure form of another inverse tetragonal structure. The cubic structure of $Mn_2PtSn$, supposedly, has the space group of $F\bar{4}3m$. As this cubic crystal structure leads to a tetragonal distortion, the new cell parameters of the tetragonal structure become $a_t = a_c/\sqrt{2}$ and $c_t = c_c$, changing the total volume of the unit cell to half; the subscripts correspond to the tetragonal and cubic structure, respectively. Upon introducing vacancies at the Mn-sites in $Mn_2PtSn$, and considering that the vacancies in the resulting compound $Mn_{1.4}PtSn$ are statistically distributed (hypothetical situation), the crystal symmetry and lattice parameters would be the same as in $Mn_2PtSn$, i.e., $I\bar{4}m2$. However, if the vacancies are ordered, it transforms the space group from $I\bar{4}m2$ to $I\bar{4}2d$, and consequently changes the lattice parameters. The new lattice parameters of $Mn_{1.4}PtSn$ become, $a_T = \sqrt{2}a_t$ and $c_T = 2c_t$, modifying the volume of the unit cell by four times; the subscripts correspond to the vacancy ordered and vacancy disordered structures, respectively. Thus, the relation between the volumes of the involved structures can be understood as, $V_T = 4V_t = 2V_c$.

In Figure 2(a), we show the crystal structure of $Mn_{1.4}PtSn$ with space groups of $I\bar{4}m2$ (hypothetical, black line) and $I\bar{4}2d$ (red line) as subcell and supercell, respectively, along [001] zonal axis. The 'a' cell parameter of the supercell, $a_T$, is equal to the diagonal of the old sub-unit cell, which is $\sqrt{2}a_t$. We employ single crystal x-ray diffraction study to understand the superstructure formation in $Mn_{1.4}PtSn$. Figure 2(c) shows the corresponding experimental single crystal XRD pattern obtained in the reciprocal space. The black square connecting the intensities of $220, \bar{2}20, 2\bar{2}0,$ and $\bar{2}\bar{2}0$ represents the subcell, while the red square connecting the intensities of $200,\ \bar{2}00,\ 020,$ and $0\bar{2}0$ represents the supercell. Figures 2(b) and (d) show the crystal structure and the experimental single crystal XRD patterns, respectively, along the [100] zonal axis. From the single crystal structure refinement, the best-fitted unit cell parameters are $a$ = 6.3651(4) Å, and $c$ = 12.2205(11) Å. We provide the structural parameters obtained from the complete single crystal refinement in Table 1. In Table 2, Wyckoff and atomic positions in the unit cell are presented along with the isotropic displacement parameters, while the anisotropic displacement of atoms are shown as Table S1 (Supplementary Information). In **Figure 2**(e), we show the observed and calculated powder XRD patterns for $Mn_{1.4}PtSn$, considering the ordered superstructure with the space group of $I\bar{4}2d$. The inset shows an expanded section of the pattern in the low angle range with *hkl* labels for the highest superstructure reflections. In the inset, the blue curve represents the simulated XRD pattern for the disordered crystal structure with a space group $I\bar{4}m2$. Upon comparing the simulated XRD patterns for both the space groups with the experimental XRD patterns, it can be clearly observed that the experimental patterns match accurately with the superstructure space group $I\bar{4}2d$, and thus rules out the possibility of disordered crystal structure with the space group $I\bar{4}m2$.

Having proven the existence of the superstructure in $Mn_{1.4}PtSn$, we proceed to understand the systematic vacancy formation in the crystal and atomic environment around different types of atoms. As stated before, the crystal structure of $Mn_{1.4}PtSn$ can be described as a superstructure of the inverse tetragonal Heusler structure ($Mn_2PtSn$) with ordered Mn-vacancies. This structure is obtained by doubling the *c*-axis, forced by an ordered arrangement of vacancies in the Pt–Mn sub structure of inverse $Mn_2PtSn$ (**Figure 1**). The created vacancies, starting from $Mn_2PtSn$, are equally distributed in Pt–Mn layers (for $Mn_{1.4}PtSn$: Pt–(0.4Mn/0.6□)), which are alternately stacked with Mn–Sn layers along the *c*-axis. Here, □ is the placeholder for a vacancy. In each Pt–(Mn/□) layer, the vacancies are arranged in such a way that they are at maximal distances from each other. In **Figure 3** we show the atomic environments as coordination shells of Mn1, Mn2, Pt, Sn and □ individually for both $Mn_2PtSn$ and $Mn_{1.4}PtSn$ in the asymmetric part of the unit cell. There is no vacancy in the atomic environment of Mn1, two in the first coordination shell of Mn2, three in the shell of Pt, two in the shell of Sn. The coordination shell of □ does not contain any vacancy.

A large magnetocrystalline anisotropy is expected in $Mn_{1.4}PtSn$ because of the tetragonal superstructure and the involvement of heavy elements Pt and Sn. We took advantage of the big single crystals and carried out directional dependent magnetic measurements of well-oriented single crystals of $Mn_{1.4}PtSn$. **Figure 4(a)** shows the magnetization vs temperature measurements in an applied external field of 0.01 T for two crystallographic axes, namely [100] and [001], in zero field cooled (ZFC) and field cooled (FC) conditions. We can clearly observe a paramagnetic to ferromagnetic transition appearing at 392 K. On decreasing the temperature further, the sharp increase in the magnetic moment, around 170 K, is attributed to a spin-reorientation transition ($T_{SR}$), below which the spin structure is non-coplanar as a result of dominating DM interaction[1,4]. The absence of any hysteresis in magnetization under warming and cooling conditions, rules out the structural origin of this transition.

We measured isothermal magnetization at 2 K and 300 K for both the directions described above. At 2 K, which is below $T_{SR}$, the magnetization along [001] saturates at 0.3 T, and the saturation magnetization is 4.7 $\mu_B$/f.u, while along [100], the saturation is rather slow and is achieved at 3 T. Thus, below $T_{SR}$, the easy axis is [001]. This magnetic behavior indicates the large magnetic anisotropy present in the compound, while the negligible coercivity shows that the compound is a soft magnet. At 300 K, which is above $T_{SR}$, [001] is still an easy axis, as shown in Figure 4(d). Unlike at 2 K, the magnetization along both [001] and [100] varies rather linearly with the magnetic field below the saturation at 300 K. We have measured the electrical resistivity of $Mn_{1.4}PtSn$ as a function of temperature (Figure 4(d)) while passing the current along [100]. The resistivity decreases with decreasing temperature, which indicates that the compound is metallic in nature. The slope changes at 392 K and 170 K correspond to $T_C$ and $T_{SR}$, respectively, which are consistent with the magnetization measurements. The values of resistivity at 300 K and 2 K are 0.15 mΩcm and 0.046 mΩcm, respectively.

In summary, we present $Mn_{1.4}PtSn$ as a design strategy to find ideal candidates for the realization of skyrmions and demonstrate the growth of its large single crystals. The method used to grow micro-twin-free single crystals reported in this work can be employed for many other compounds. The $D_{2d}$ symmetry of the superstructure form of tetragonal inverse Heusler $Mn_{1.4}PtSn$ is essential for stabilizing the antiskyrmions. Other Mn-rich Heusler compounds incorporating 4d and 5d transition metals and heavy main group elements are expected to uncover interesting non-collinear spin behavior, including skyrmions. Another aspect of interest is the occurrence of ferromagnetism in $Mn_{1.4}PtSn$, which we believe to originate from the Mn-vacancy ordering. The absence of such ordering in $Mn_2RhSn$ leads to ferrimagnetic interactions.

**EXPERIMENTAL SECTION**

**Single crystal growth**

Single crystals of $Mn_{1.4}PtSn$ were grown by flux-growth method using Sn as a flux. Highly purified Mn (99.999%) and Pt (99.999%) were cut into very small pieces and weighed in 3:1 molar ratio, with a total weight 0.75 g. This stoichiometric portion was loaded in a dried alumina crucible together with 10 g of Sn. Subsequently, the alumina crucible was sealed in a quartz tube under 0.2 bar of argon pressure. The quartz ampule was put in the box furnace and heated to 1323 K at a rate of 200 K/h. The whole content was maintained at this temperature for 24 h, for homogeneity. Initially, the furnace temperature was rapidly decreased to 923 K, after which slow cooling to 723 K occurred, at a rate of 2 K/h, for the crystal growth. At 723 K, the tube was kept again for 24 h before removing the extra flux by centrifugation. From this procedure, silvery single crystals with 1-2 mm sizes, were obtained; the crystals are air and moisture stable and no decomposition was observed even after several months. Under the same experimental conditions, we grew different batches of crystals, which indicated the reproducibility of the method. Several crystals were separated for further characterizations.

**Characterization**

Single crystal X-ray diffraction data were collected by a Rigaku AFC7 diffractometer with a Saturn 724+ CCD detector, and refined by SHELXL.[20,21] Laue diffraction patterns were simulated with the OrientExpress software. The compositions of the compound were characterized by SEM-EDAX and chemical analysis. The composition obtained from the EDAX analysis is 41% Mn, 30% Pt, and 29% Sn and the composition obtained from chemical analysis is $Mn_{1.43}PtSn_{1.01}$, which are very close to the compositions obtained from single crystal X-ray analysis ($Mn_{41.86}Pt_{29.07}Sn_{29.07} \approx Mn_{1.44}PtSn$). Differential scanning calorimetry (DSC) of the sample (~56 mg) was performed using the NETZSCH DSC 404C instrument. Magnetic properties were measured in SQUID VSM (Quantum Design). The magnetic measurements were performed in the temperature range of 2-400 K, and the magnetic field range $(-)7-(+)7$ T. The electrical resistivity was measured from 2-400 K in a Physical Property Measurement System.

**ASSOCIATED CONTENT**

The Supporting Information is available free of charge on the ACS Publications website at DOI:
Differential scanning calorimetry (DSC) of $Mn_{1.4}PtSn$; Polarized light optical micrograph of $Mn_{1.4}PtSn$; Single crystal figure of $Mn_{1.4}PtSn$; Laue patterns in different crystallographic directions; Composition by chemical analysis; Magnetization vs temperature at various applied magnetic field; Anisotropic displacement parameters and bond lengths.

## AUTHOR INFORMATION

### Corresponding Author

* Prof. Claudia Felser; Claudia.Felser@cpfs.mpg.de

### Present Addresses

† Max Planck Institute for Chemical Physics of Solids 01187 Dresden, Germany

### Notes

The authors declare no competing financial interest.

## ACKNOWLEDGEMENTS

This work was financially supported by the ERC Advanced Grant 742068 "TOPMAT."

**Figures:**

Figure 1.

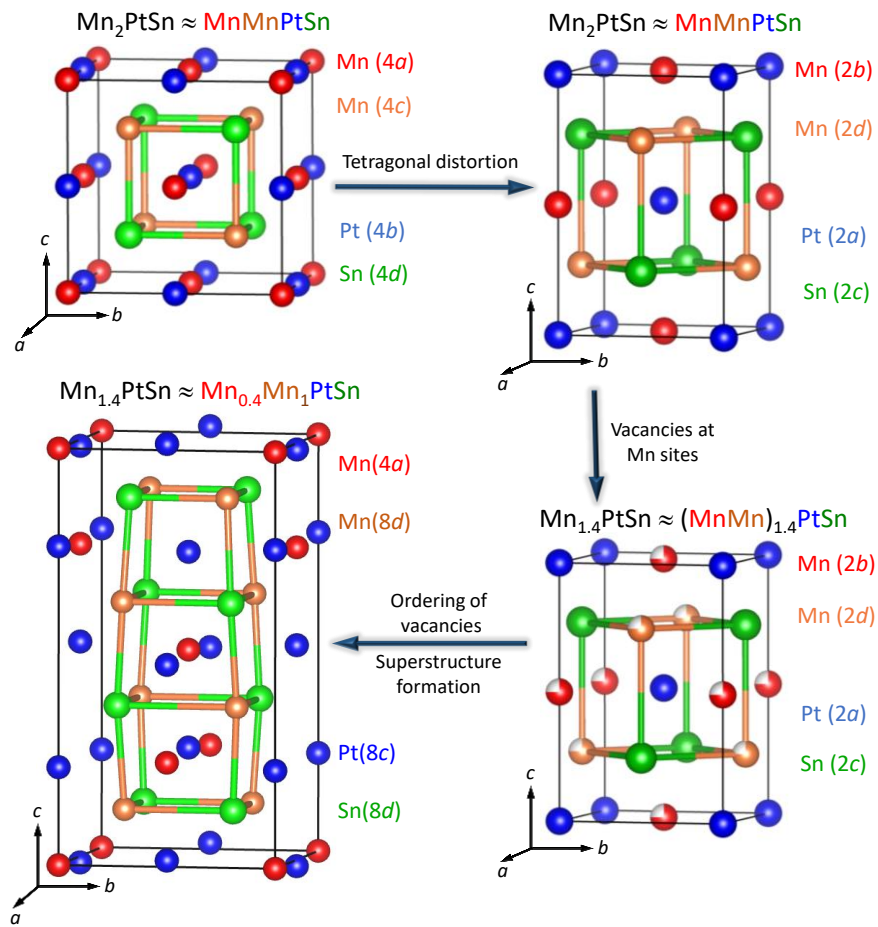

**Figure 1.** Schematic evolution of the superstructure induced by ordering of the Mn-vacancies in $Mn_{1.4}PtSn$. The Mn-Pt sub structure is shown by black solid lines whereas the Mn-Sn sub structure is shown by green-orange solid lines.

Figure 2.

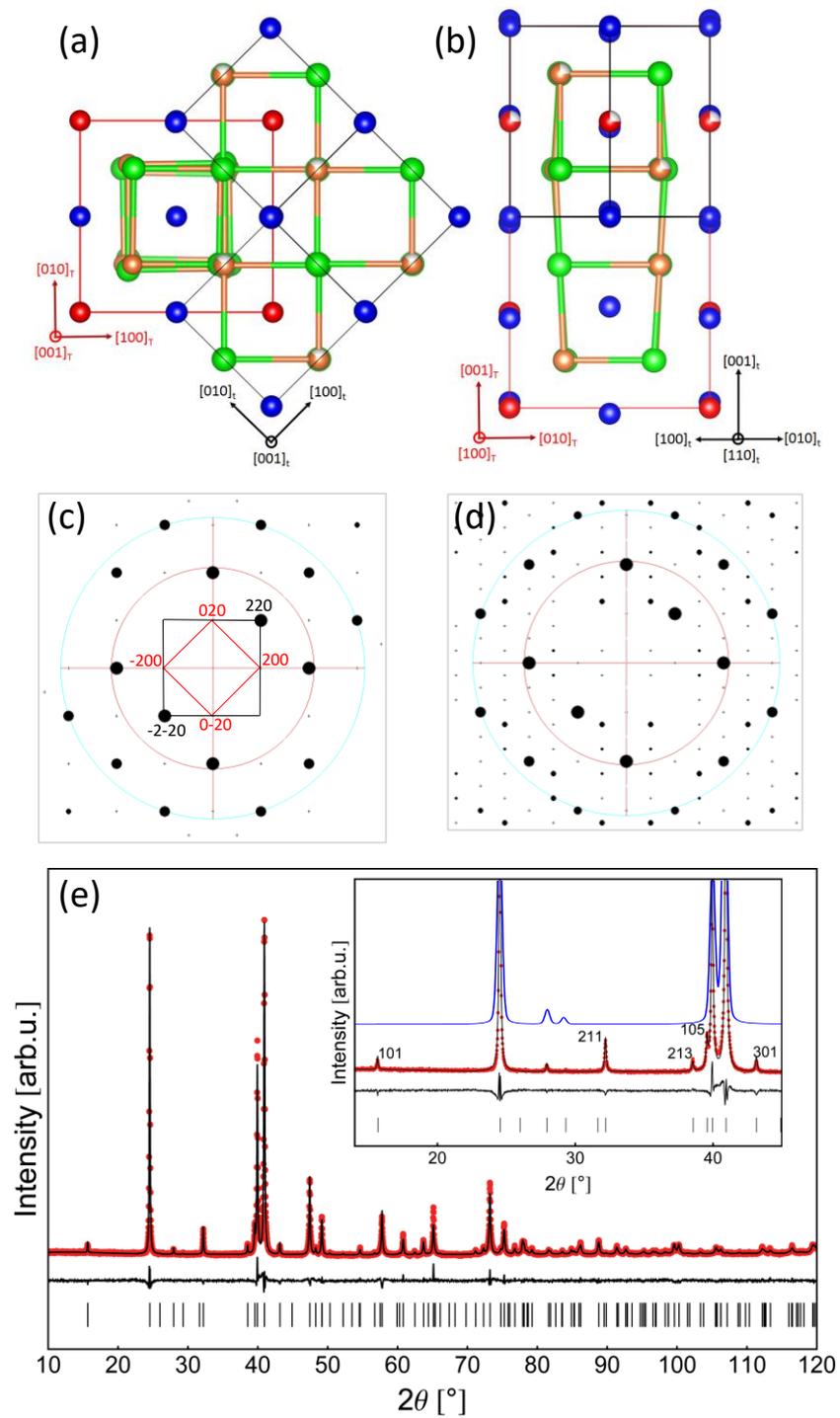

**Figure 2.** (a), (b), and (c), (d) Show real and inverse space patterns for the zone axis [001] and [100], respectively. (e) Observed and calculated powder X-ray diffraction patterns. The inset shows the pattern with *hkl* labels for the higher superstructure reflections. For comparison, we show the simulated pattern corresponding to the disordered crystal structure in space group $I\bar{4}m2$ with missing superstructure reflections.

Figure 3.

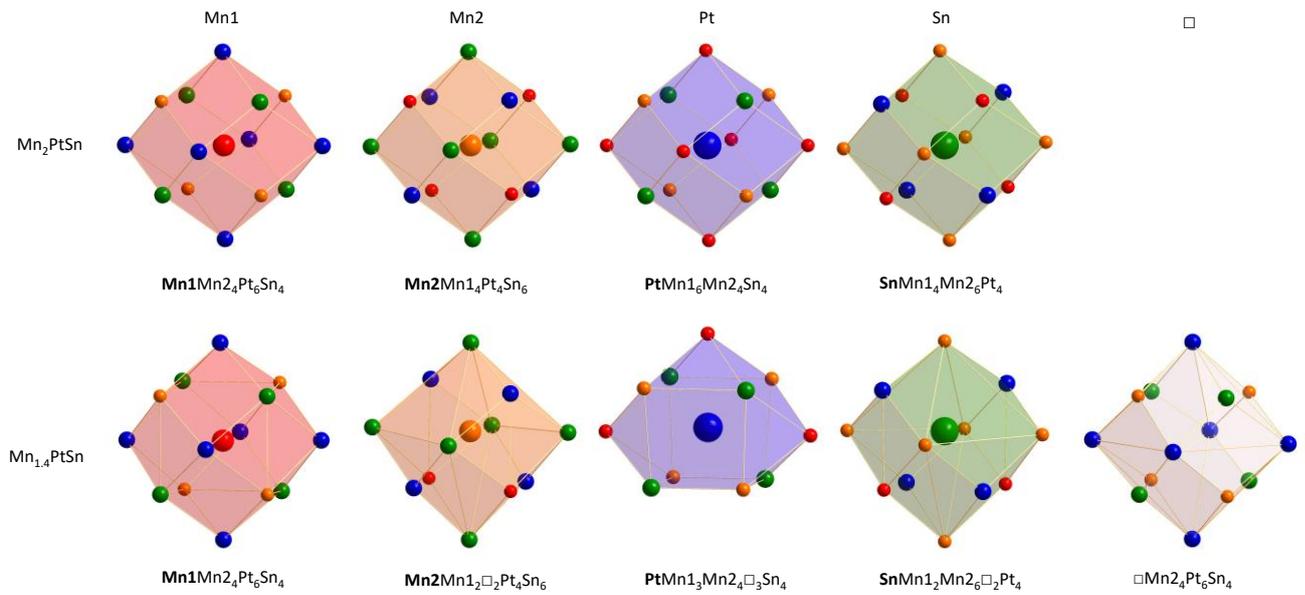

**Figure 3.** Atomic environments of the atoms in the asymmetric part of the unit cell, and vacancies for (first row) tetragonal inverse $Mn_2PtSn$ and (second row) $Mn_{1.4}PtSn$.

Figure 4.

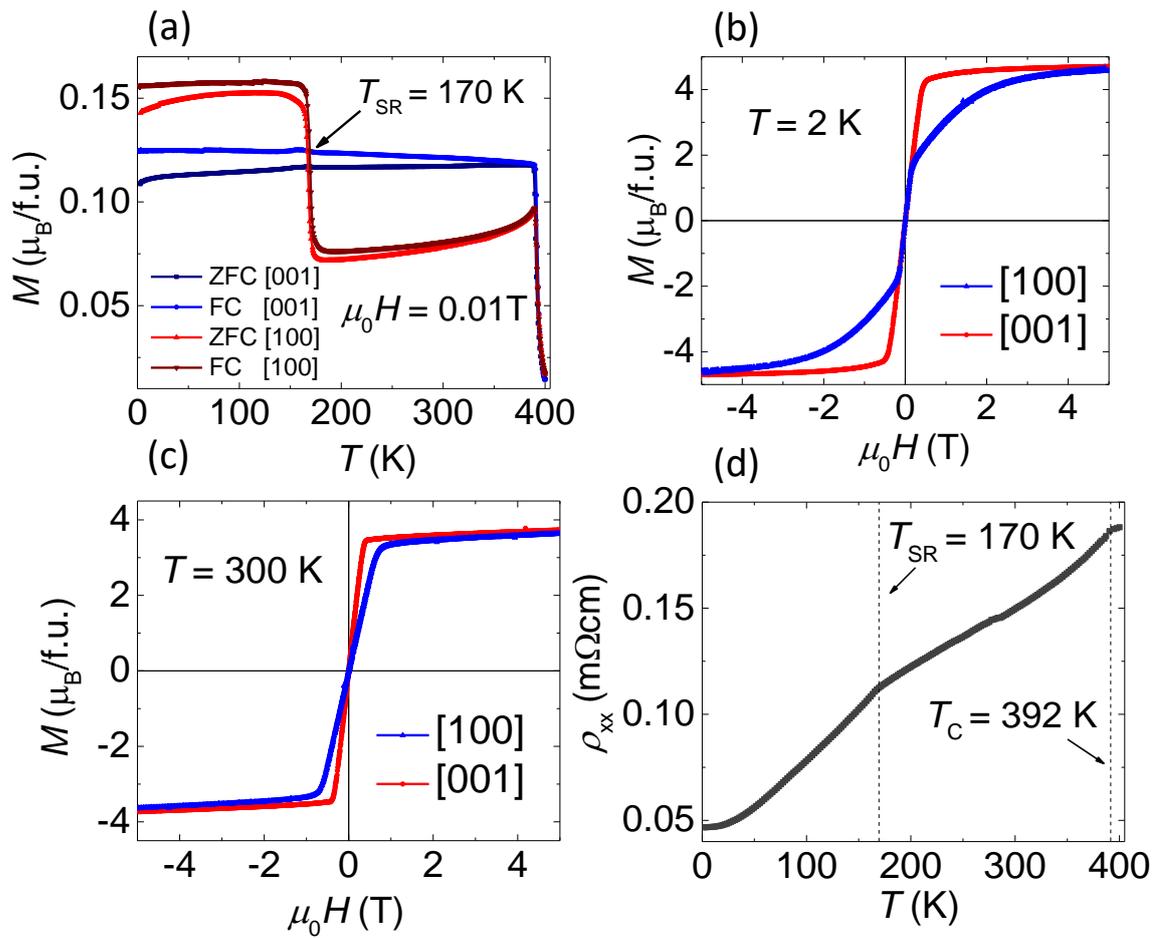

**Figure 4.** (a) Magnetization vs temperature at 0.01 T in different directions, isothermal magnetization at (b) 2 K and (c) 300 K. (d) Temperature dependent resistivity from 2 to 400 K for 0 T field.

**Tables**

Table 1.

Table 1. Single-crystal X-ray diffraction refinement details for Mn$_{1.4}$PtSn.

| Compound | Mn$_{1.44}$PtSn |
|---|---|
| Formula weight | 392.96 |
| Temperature [K], wavelength [Å] | 295, 0.71703 (Mo Kα) |
| Space group | $I\bar{4}2d$ |
| Unit cell parameters [Å] | $a$ = 6.3651(4), $c$ = 12.2205(11) |
| Volume [Å$^3$] | 495.11(8) |
| reflections collected, independent | 5060, 692 (R$_{int}$=0.039) |
| theta range [°], completeness | 3.609 - 36.964, 100 % |
| data/restraints/parameters | 632/0/20 |
| GoF | 0.954 |
| $R$ indices (all data)[a] | R1= 0.0203, wR2= 0.0377 |

[a] $R = \sum ||F_0| - |F_c||/\sum |F_0|$, $wR2 = (\sum[w(|F_0|^2 - |F_c|^2)^2]/\sum[w(|F_0|^4)]^{1/2}]$ and $w = 1/[\sigma^2(|F_0|^2) + (0.0154\ P)^2 + 5.5965\ P]$, where $P = (F_0^2 + 2\ F_c^2)/3$.

Table 2.

Table 2. Fractional Atomic Coordinates and Isotropic Equivalent Atomic Displacement Parameters for Mn$_{1.4}$PtSn

| Atoms | Wyckoff position | $x$ | $y$ | $z$ | $U_{eq}$ (Å$^2$) |
|---|---|---|---|---|---|
| Mn1 | 4$a$ | 0 | 0 | 0 | 0.012 |
| Mn2 | 8$d$ | 1/4 | 0.22820 | 3/8 | 0.01 |
| Pt | 8$c$ | 0 | 1/2 | 0.48488 | 0.011 |
| Sn | 8$d$ | 0.27875 | 1/4 | 1/8 | 0.012 |

# Table of Contents

We present Mn$_{1.4}$PtSn as a design principle to search for skyrmions magnetic materials. The large micro-twin-free single crystals allow us to discover first superstructure formation in the large family of Heusler compounds.

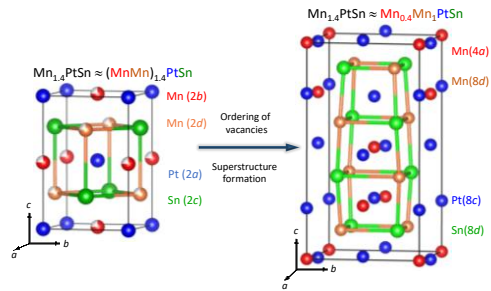

*Praveen Vir, Nitesh Kumar, Horst Borrmann, Bayardulam Jamijansuren, Guido Kreiner, Chandra Shekhar, and Claudia Felser\**

*1 – 5*

**Tetragonal superstructure of antiskyrmion hosting Heusler compound Mn$_{1.4}$PtSn**

# Supplementary Information

## Tetragonal superstructure and properties of antiskyrmion hosting Heusler compound Mn$_{1.4}$PtSn

Praveen Vir, Nitesh Kumar, Horst Borrmann, Bayardulam Jamijansuren, Guido Kreiner, Chandra Shekhar, and Claudia Felser[*]

*Max Planck Institute for Chemical Physics of Solids, 01187 Dresden, Germany*

**Differential Scanning Calorimetry (DSC) and polarized light micrograph**

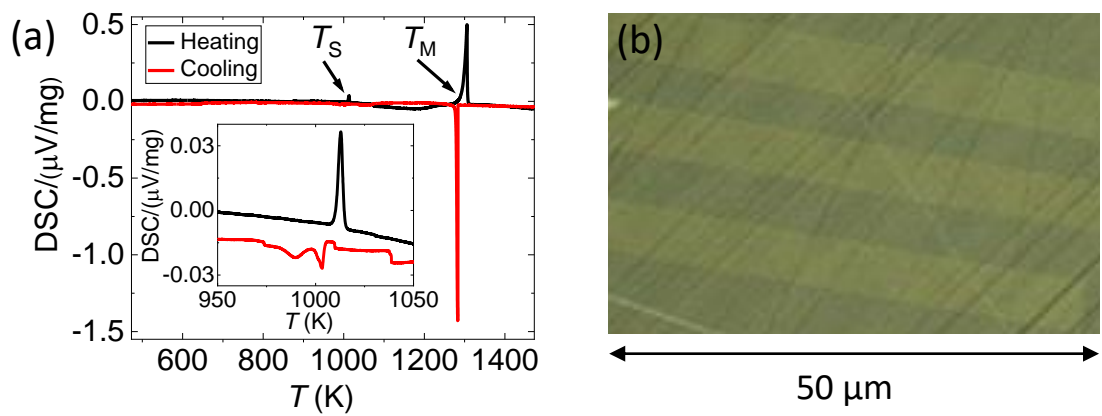

**Figure S1.** (a) DSC measurement from 473 to 1473 K: inset is enlarged part of region having peaks corresponding to structural transition (b) Crystal having microtwinned structure obtained from Polarized light microscope.

Our DSC analysis reveals a (Figure S1(a)) structural transition in Mn$_{1.4}$PtSn at 1013 K. This transition corresponds to a high-temperature cubic phase to low-temperature tetragonal phase. Such structural transformation is rather a common feature and it is also observed in other non-centrosymmetric tetragonal Heusler compounds[1–3]. This structural transition is termed as the martensitic transition. The high-temperature cubic phase is often referred to as austenite phase while the low-temperature tetragonal phase is called martensite phase. Usually, this type of structural transition is diffusion-less therefore while coming from high-temperature phase, it always leads to the formation of micro-twinned grains in the compound. The detail of kind of structural transition is very well studied in magnetic shape-memory compounds and magnetocaloric compounds[4,5]

The hysteresis observed in the heating and cooling curve for both these peaks indicates that the transformations are of the first order. The absence of any other peak around the melting point indicates that the compound melts congruently and it does not decompose to any other phase.

In Figure S1(b), we show polarized light optical micrograph of the surface of a crystal grown above martensitic transition temperature. There exist regions with different contrasts which correspond to different orientations. Upon further analysis by transmission electron microscopy, it was found out that these grains are crystallographically perpendicularly oriented.

**Crystal**

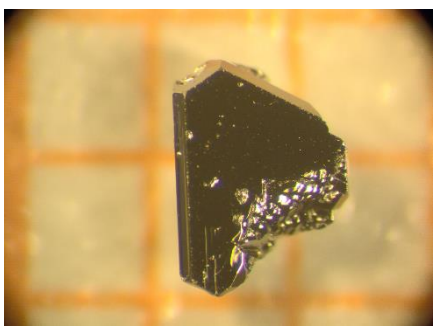

**Figure S2.** As grown single crystal from flux-method on mm-scale grid paper.

**Composition by chemical analysis**

We have performed chemical analysis on crystals of $Mn_{1.4}PtSn$ in order to know the composition. The instrument used for this process was ICPOES S100 SUDV (Agilent) and the digestion process was done with Turbowave (MLS). The acidic solution, used for dissolving crystals was of 3mL HCl and 0.5mL $HNO_3$. Four times independent, 5 mg of crystals were used to get the standard deviation in the composition. The composition obtained is listed in the following table,

| Elements | Composition (wt %) ± deviation | Composition (molar ratio) ± deviation |
|---|---|---|
| Mn | 20.11±0.13 | 1.426±0.011 |
| Pt | 50.08±0.24 | 1±0.007 |
| Sn | 31.05±0.09 | 1.019±0.006 |

The obtained composition ($Mn_{1.43}PtSn$) from the chemical analysis is very close to the composition ($Mn_{1.44}PtSn$) obtained from single crystal XRD.

**Laue patterns**

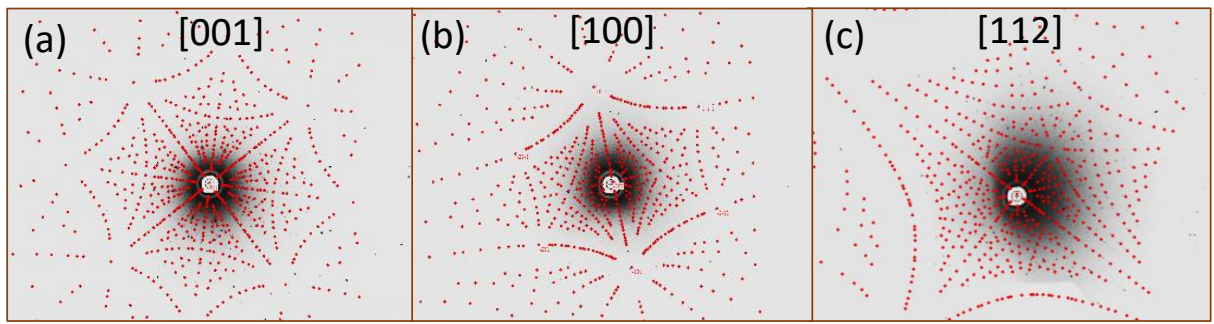

**Figure S3.** Laue patterns for the crystallographic direction (a) [001] (b) [100], and (c) [112].

In order to proceed for measuring directional dependent physical properties, we oriented single crystals of $Mn_{1.4}PtSn$ by Laue diffraction method. In every growth process batch, it was found out that the crystals have two types of crystal facets i.e. rectangular and hexagonal. The rectangular facets can be either (100) or (001) while the hexagonal ones are always (112). Figure S3(a), (b) and (c) show the Laue diffraction patterns along with the superimposed simulated patterns along [100], [001] and [112] crystal directions. Laue patterns along [100] show two-fold rotation while along [001] it shows four-fold rotation which implies the presence of tetragonal symmetry in the crystal. Laue pattern along [112] direction which does not contain any rotational symmetry as expected from the tetragonal structure.

**Magnetization vs temperature**

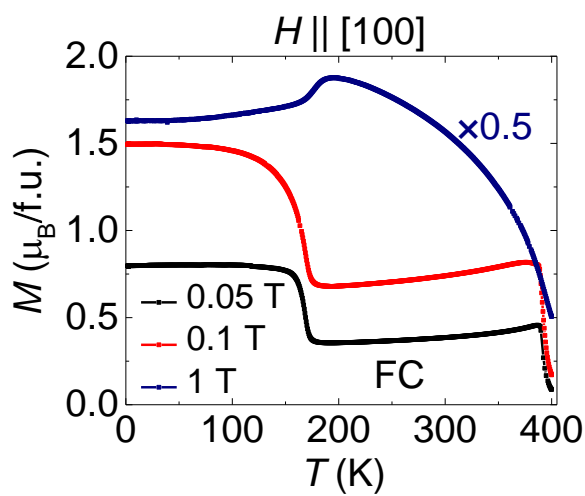

**Figure S4.** $M$ vs $T$ at the external field of 0.05 T, 0.1 T, and 1 T.

To check the dependency of these transitions on the magnetic field, we extended our magnetic measurements in the higher field along [100]. Figure S4 shows the magnetization as a function of temperature in field cooled condition at 0.05 T, 0.1 T, and 1 T. As it can be seen from the plot, $T_C$, and $T_{SR}$ are almost unaffected with the applied magnetic field. For 0.05 and 0.1 T, the moment values for temperatures below spin-reorientation transition is higher than for temperatures higher spin-reorientation transition.

**Tables**

**Table S1:** Anisotropic displacement parameters (in Å$^2$) for Mn$_{1.44}$PtSn at 295(2) K with estimated standard deviations in parenthesis

| Label | $U_{11}$ | $U_{22}$ | $U_{33}$ | $U_{12}$ | $U_{13}$ | $U_{23}$ |
|---|---|---|---|---|---|---|
| Mn(1) | 0.0153(8) | 0.0153(8) | 0.0060(7) | 0 | 0 | 0 |
| Mn(2) | 0.0089(7) | 0.0119(9) | 0.0075(5) | 0 | 0 | -0.0014(10) |
| Pt(1) | 0.00968(16) | 0.01545(17) | 0.00811(12) | 0.0006(2) | 0 | 0 |
| Sn(1) | 0.0145(3) | 0.0142(4) | 0.0079(2) | 0 | 0 | 0.0003(5) |

$^a$The anisotropic displacement factor exponent takes the form: $-2\pi^2$ [$h^2a^{*2}U_{11}$ + … + 2hka* b* $U_{12}$].

**Table S2:** Bond lengths (Å) for Mn$_{1.44}$PtSn at 295(2) K with estimated standard deviations in parentheses

| Label | distances | Label | distances |
|---|---|---|---|
| Pt(1)- Mn(1) | 2.7068(13) ×2 | Pt(1)- Mn(2) | 3.1879(2) ×2 |
| Pt(1)- Sn(1) | 2.7288(6) ×2 | Sn(1)- Mn(2) | 2.8310(7) ×2 |
| Pt(1)- Sn(1) | 2.7358(7) ×2 | Sn(1)- Mn(1) | 2.8610(16) |
| Pt(1)- Mn(1) | 2.7523(10) ×2 | Sn(1)- Mn(1) | 3.0637(3) ×2 |
| Pt(1)- Mn(2) | 2.8705(4) | Sn(1)- Mn(1) | 3.1829(2) ×2 |
| Mn(1)- Mn(2) | 2.8030(12) ×2 | | |